\documentclass{article}

\usepackage{epsf,a4wide,cite}

\begin{document}
\hbox{}
\nopagebreak
\vspace{-3cm}
\begin{flushright}
{\sc LBNL-42095} \\
{\sc OUTP-98 55P} \\
{\sc Cavendish-HEP-98/12}\\
\end{flushright}

\vspace{1in}

\begin{center}
{\Large \bf Unitarization of Gluon Distribution in the Doubly
Logarithmic Regime at High Density}
\vspace{0.1in}

\vspace{0.5in}
{\large  Jamal Jalilian-Marian$^1$, Alex Kovner$^2$, Andrei Leonidov$^3$
and Heribert Weigert$^4$}\\
{\small
$^1${\it Nuclear Theory Group, Nuclear Science Division, LBNL, 
Berkeley, CA , USA}\\
$^2${\it Theoretical Physics, Oxford University, 1 Keble road, Oxford,
OX1 3NP, UK}\\
$^3${\it Theoretical Physics Department, P.N.~Lebedev Physics Institute,\\
117924 Leninsky pr. 53  Moscow, Russia}\\
and
{\it Theoretical Physics Institute, University of Minnesota, 116
Church st. S.E., Minneapolis, MN 55455, USA}\\
$^4${\it University of Cambridge, Cavendish Laboratory, HEP, Madingley Road,
Cambridge CB3 0HE UK}
}

\vspace{0.5in}

{\bfseries\sc  Abstract}\\
 
\vspace{.2in}
\begin{minipage}{.9\textwidth}
  We analyze the general nonlinear evolution equations for multi gluon
  correlators derived in \cite{moap} by restricting ourselves to a
  double logarithmic region.  In this region our evolution equation
  becomes local in transverse momentum space and amenable to
  simple analysis. It provides a complete nonlinear
  generalization of the GLR equation. We find that the
  full double log evolution at high density becomes strikingly
  different from its linear doubly logarithmic DGLAP counterpart. An
  effective mass is induced by the nonlinear corrections which at high
  densities slows down the evolution considerably.  We show that at
  small values of impact parameter the gluonic density grows as a
  logarithm of energy.  At higher values of impact parameter the
  growth is faster, since the density of gluons is lower and
  nonlinearities are less important.
\end{minipage}
\end{center}

\vfill

\newpage

\section {Introduction}

One of the most important theoretical problems in the high energy
hadronic physics is the understanding of high density nonlinear
effects which may lead to unitarization of perturbatively generated
growth of the hadronic cross sections at very high energy. The
simplest process in which one hopes to see these effects is deep
inelastic scattering. Standard linear evolution~\cite{dglap}, which is
phenomenologically so successful at present energies predicts a
perturbatively generated growth of the total cross section faster than
a power of a logarithm. This contradicts the expectation based on the
unitarity constraint that the growth should not be faster than the
second power of the logarithm of energy\footnote{ There is a certain
  caveat here, since in a strict sense there is no proof that the DIS
  cross section has to unitarize in the same way as purely hadronic
  cross sections.  Still one believes based on physical arguments that
  the growth of the DIS cross section at high $Q^2$ should not exceed
  a power of a logarithm~\cite{levin1,frankfurt1}.}.  It is natural to
assume that the effects that are left out of the linear evolution are
the ones that are ultimately responsible for the unitarization
effects.  In particular the suspicion falls on possible screening
effects due to finite partonic density. At high energy this density
grows and the emission of high transverse momentum partons which are
scattered by the probe should be inhibited relative to the low density
situation \footnote{Another possibility is that even at high $Q^2$ at
  high enough energy the scattering becomes nonperturbative due to
  large contributions from the small transverse momentum region. This
  situation then is entirely outside the reach of the methods of
  perturbative QCD. It has been however convincingly argued that even
  if this is the case asymptotically, there are many physically
  interesting situations where the main contribution to the cross
  section come from the perturbative
  region~\cite{muller1,levin2,hebbecker1}. In these cases the
  perturbative nonlinearities must play a crucial role.}.  These
nonlinear effects are not taken into account by the linear evolution
equations.  Recently it has been argued that already the presently
available data on DIS can not be reasonably explained without taking
into account the nonlinear effects in QCD evolution \cite{elevin1}.

Possible nonlinear generalizations of perturbative evolution have been
discussed many times in literature starting with the famous GLR model
\cite{glr}.  A fully nonlinear evolution equation based on a
particular model for the gluon source was derived by Mueller in
\cite{muller2}, which inspired much of the later research in this
field.  More recently generalized nonlinear evolution equations were
discussed by Levin and Laenen \cite{levin3} and Ayala, Gay-Ducati and
Levin \cite{levin4}.  All these approaches are based to some extent on
a physically motivated ansatz and it is important to understand the
impact of nonlinearities in a more general framework.

In a recent series of papers \cite{rg,bfkl,nonlin,moap} we have
developed an approach to the evolution of dense partonic systems
within the framework of the Wilson renormalization group\footnote{Let
  us note that there exists a number of approaches to small-x physics
  based either directly on constructing the QCD effective action at
  small $x$ by combining the reggeon and usual gluon terms
  \cite{lipatov} (see also \cite{kp}), or on the generalization of the
  operator product expansion \cite{ib}. It is important to understand
  the relation between these different approaches and the approach
  described in this paper. At the moment this relation is not
  completely clear to us.}. The method is based on earlier work by
McLerran, Venugopalan and Ayala, Jalilian-Marian, McLerran and
Venugopalan \cite{mv} in which the idea of representing fast partons
by static color charge density was developed.  This approach
results in a nonlinear functional evolution equation for the
generating functional of the color charge density correlators, which
is valid to leading order in $\alpha_s$ at densities which
parametrically do not exceed $1/\alpha_s$. This is indeed the region
of density which is interesting for the perturbative unitarization
effects.  The equation is fairly complicated since it only requires
ordering in longitudinal momenta during evolution and puts no
constraint of any kind on the ordering of transverse momenta. In fact,
in the low density limit it reduces to the BFKL equation for the two
point correlation function \cite{bfkl} and a simple truncation of it
reproduces the complete BKP hierarchy \cite{prep}.

The aim of the present paper is twofold.  First we are going to show
how by a simple transformation this evolution equation is converted
into a set of equations for the evolution of the correlators of the
chromoelectric field rather than the color charge density.  This form
is perhaps more advantageous, since these correlators are more
directly related to observable quantities and for example the DGLAP
evolution operates directly with them.  Our second goal is to consider
these evolution equations in the doubly logarithmic regime.  That is
to say we will work in the approximation where strict transverse
momentum ordering is imposed on the evolution. In our framework this
corresponds to the leading order in the expansion in powers of
transverse derivatives.  It turns out that the evolution equations
simplify tremendously in this limit and become much more tractable.

The main feature of the full nonlinear evolution is the appearance of
a dynamical mass. This mass is 
induced by quantum corrections to the QCD evolution
at finite density and is proportional to the square of the chromoelectric
field. Its effect is to slow 
down the evolution relatively to the standard
perturbative case. As a consequence the gluon density in the
small impact parameter region at high density grows with energy only
logarithmically as opposed to the $\exp\sqrt{\ln s}$ type of growth at
low density.  To construct an explicit solvable model we approximate
the generating functional for the chromoelectric field by a Gaussian.
In this Gaussian approximation the only independent quantity is the
distribution function itself, that is the two point correlation
function of the chromoelectric field. All higher correlators have
simple factorization properties in terms of the two point function.
This model leads to a simple closed nonlinear equation for the
distribution function which is qualitatively quite similar to the
equation considered in \cite{levin4}.

The paper is structured as follows. In Sec. 2 we briefly review the
general structure of the nonlinear evolution equation and show how to
rewrite it directly in terms of correlators of the chromoelectric
field.  In Sec. 3 we explain how the double logarithmic limit
arises in the present framework and derive the explicit evolution
equation in this limit.  Sec. 4 is devoted to analysis of the the
resulting equation and the derivation of the simple Gaussian model.
Finally Sec. 5 contains a brief discussion.

\section{Evolution equations for the correlators of the 
chromoelectric field.}

First let us briefly recall the framework and the results of
\cite{moap}.  In this approach the averages of gluonic observables in
a hadron are calculated via the following path integral
\begin{eqnarray}
\label{action}
&&<O[A]>=\int D\rho DA_\mu
O[A]\exp\left\{ -\int d^2 x_\perp F[\rho ^a(x_\perp)] 
\right. \\ && \left.
-i\int d^4 x {1\over 4} F_a^{\mu\nu}F^a_{\mu\nu}
 -{{1}\over{N_c}} \int d^2 x_\perp dx^-
\delta (x^-)
\rho^{a}(x_\perp) {\rm tr}T_a 
W_{-\infty,\infty} [A^-](x^-,x_\perp)\right\}
\nonumber 
\end{eqnarray}
where $W$ is the Wilson line in the adjoint representation along the
$x^+$ axis.  The hadron is represented by an ensemble of color charges
localized in the plane $x^-=0$ with the (integrated across $x^-$)
color charge density $\rho(x_\perp)$.  The statistical weight of a
configuration $\rho(x_\perp)$ is
\begin{equation}
Z=\exp \{-F[\rho]\}
\end{equation}
In the tree level approximation (in the light cone gauge $A^+=0$) the
chromoelectric field is determined by the color charge density through
the equations
\begin{equation}
F^{+i}={1\over g}\delta(x^-)\alpha_i(x_\perp)
\label{chrom}
\end{equation}
and the two dimensional vector potential $\alpha_i(x_\perp)$ is "pure
gauge" and is related to the color charge density by
\begin{eqnarray}
&&\partial_i\alpha^a_j-
\partial_j\alpha^b_i-f^{abc}\alpha^b_i\alpha^c_j=0\nonumber \\
&&\partial_i\alpha^a_i=-\rho^a
\label{sol}
\end{eqnarray}

Integrating out the high longitudinal momentum modes of the vector
potential generates the renormalization group equation, which has the
form of the evolution equation for the statistical weight $Z$
\cite{nonlin,moap} \footnote{All the functions in the rest of this
  paper depend only on transverse coordinates.  To simplify notation
  we largely drop the subscript $\perp$ in the following.}
\begin{equation}
{d\over d\zeta}Z= \alpha_s \left\{{1\over 2}{\delta^2
\over\delta\rho(u)\delta\rho(v)}\left[Z\chi(u, v) \right] -{\delta
\over\delta\rho(u)}\left[Z\sigma(u)\right]\right\}
\label{final}
\end{equation}
In the compact notation used in Eq.~(\ref{final}), both $u$ and $v$
stand for pairs of color index and transverse coordinate, with
summation and integration over repeated occurrences implied.  The
evolution in this equation is with respect to the rapidity $\zeta$,
related to the Feynman $x$ by
\begin{equation}
\zeta=\ln 1/x
\end{equation}
Technically it arises as a variation of $Z$ with the cutoff imposed on
the longitudinal momentum of the fields $A_\mu$.  The quantities
$\chi[\rho]$ and $\sigma[\rho]$ have the meaning of the mean
fluctuation and the average value of the extra charge density induced
by the high longitudinal momentum modes of $A_\mu$. They are
functionals of the external charge density $\rho$. The explicit
expressions have been given in \cite{moap}. For the purpose of the
general discussion in this section we do not need their explicit form.
Eq.~(\ref{final}) can be written directly as evolution equation for the
correlators of the charge density. Multiplying Eq.~(\ref{final}) by
$\rho(x_1)...\rho(x_n)$ and integrating over $\rho$ yields
\begin{eqnarray}
\lefteqn{
{d\over d\zeta}<\rho(x_1)...\rho(x_n)>=
}
\\ &&\alpha_s
\left[\sum_{0<m<k<n+1}<\rho(x_1)...\rho(x_{m-1})\rho(x_{m+1})... 
\rho(x_{k-1})\rho(x_{k+1})...
\rho(x_n)\chi(x_m, x_k)>
\right. \nonumber\\ && \left. \hspace{1cm}
+\sum_{0<l<n+1}<\rho(x_1)...\rho(x_{l-1})\rho(x_{l+1})
...\rho(x_n)\sigma(x_l)>\right]
\nonumber 
\label{correl}
\end{eqnarray}
In particular, taking $n=2$ we obtain the evolution equation for the
two point function
\begin{equation}
{d\over d\zeta}<\rho(x)\rho(y)>=
\alpha_s\left\{<\chi(x, y)+\rho(x)\sigma(y)+\rho(y)\sigma(x)>\right\}
\label{prop}
\end{equation}

This set of equations for the correlators of the color charge density
completely specifies the evolution of the hadronic ensemble as one
moves to higher energies (or lower values of $x$).

Our aim in this section is to convert Eq.~(\ref{correl}) into a set of
equations which determine directly the evolution of the multi particle
correlators of the chromoelectric field $\alpha_i$ \footnote{The
  chromoelectric field strength $F^{i+}$ is proportional to $\alpha_i$
  as per Eq.~(\ref{chrom}). In what follows we will therefore refer to
  $\alpha_i$ as to the field strength.}.  For this purpose we multiply
Eq.~(\ref{final}) not by $\rho(x_1)...\rho(x_n)$ but rather by
$\alpha^{a_1}_{i_1}(x_1)...\alpha^{a_n}_{i_n}(x_n)$ and again
integrate over $\rho$.  This results in a set of equations analogous
to Eq.~(\ref{correl})
\begin{eqnarray}
\lefteqn{
{d\over d\zeta}<\alpha^{a_1}_{i_1}(x_1)...\alpha^{a_n}_{i_n}(x_n)>
}
\\ & = &
\alpha_s
\bigg[\sum_{0<l<n+1}
<\alpha^{a_1}_{i_1}(x_1)...\alpha^{a_{l-1}}_{i_{l-1}}(x_{l-1})
\alpha^{a_{l+1}}_{i_{l+1}}(x_{l+1})...
\alpha^{a_n}_{i_n}(x_n)\sigma^{a_l}_{i_l}(x_l)>+ 
\nonumber\\  && \hspace{1cm}
\sum_{0<m<k<n+1}<\alpha^{a_1}_{i_1}(x_1)...\alpha^{a_{m-1}}_{i_{m-1}}
(x_{m-1})\alpha^{a_{m+1}}_{i_{m+1}}(x_{m+1})...
\nonumber \\ &&  \hspace{3cm} \times
\alpha^{a_{k-1}}_{i_{k-1}}(x_{k-1})
\alpha^{a_{k+1}}_{i_{k+1}}(x_{k+1})...
\alpha^{a_n}_{i_n}(x_n)\chi^{a_ma_k}_{i_mi_k}(x_m, x_k)>
\bigg]\nonumber
\label{correlf}
\end{eqnarray}
Here we have defined
\begin{eqnarray}
&&\chi^{ab}_{ij}(x,y)>=r^{ac}_i(x,u)\chi^{cd}(u,v)r^{\dagger db}_j(v,y)\\
&&\sigma^{a}_{i}(x)=r^{ab}_i(x,u)\sigma^b(u)+p^{abc}_i(x,u,v)\chi^{bc}(u,v)
\nonumber
\label{indfield}
\end{eqnarray}
with
\begin{eqnarray}
&&r^{ab}_i(x,y)=\frac{\delta\alpha^a_i(x)}{\delta\rho^b(y)}\\
&&p^{abc}_i(x,y,z)=\frac{\delta^2\alpha^a_i(x)}{\delta\rho^b(y)
\delta\rho^c(z)}\nonumber
\label{rp}
\end{eqnarray}
In Eqs.~(\ref{indfield}),~(\ref{rp}) the repeated indices are summed
over.

We now have to express the quantities $r$ and $p$ explicitly in terms
of the field $\alpha_i$. Once this is done the equations
Eq.~(\ref{correlf}) become explicit equations for the evolution of the
correlators of the field $\alpha_i$, since the charge density $\rho$ is
already known in terms of $\alpha_i$ by virtue of Eq.~(\ref{sol}).

To find $r^{ab}_i$ we differentiate Eq.~(\ref{sol}) with respect to
$\rho$ and get
\begin{eqnarray}
&&\partial_ir^{ab}_i=-\delta^{ab}\\
&&\epsilon_{ij}\partial_ir^{ab}_j-2f^{acd}\epsilon_{ij}
\alpha^{c}_ir^{db}_j=0\nonumber
\label{solr}
\end{eqnarray}
This set of equations is easily solved by decomposing $r_i$ into the
longitudinal and the transverse parts according to
\begin{equation}
r^{ab}_i=\partial_il^{ab}+\epsilon_{ij}\partial_jt^{ab}
\end{equation}
with the following result
\begin{equation}
r^{ab}_i(x,y)=-<x|\left\{\partial_i\delta^{ab}
+\epsilon_{ij}\partial_j
\left[\frac{1}{D\partial}\right]^{ac}
\epsilon_{kl}\alpha^{cb}_k\partial_l\right\}
\frac{1}{\partial^{2}}|y>
=-<x|\left[D_i\frac{1}{\partial D}\right]^{ab}|y>
\label{r}
\end{equation}
Here $<x|O|y>$ denotes a configuration space matrix element in the
usual sense. Scalar products with respect to space time indices here
and below refer only to transverse indices, so that for example
$\partial D = \partial_i D_i$ with $i=1,2$. For
convenience we have also defined 
\begin{eqnarray}
\label{def}
\alpha^{ab}_i&=&f^{abc}\alpha_i^c\\
D^{ab}_i&=&\partial_i\delta^{ab}+\alpha_i^{ab}\, .
\nonumber
\end{eqnarray}

Now differentiating Eq.~(\ref{solr}) once again with respect to $\rho$
we can solve for $p$:
\begin{equation}
\label{p}
p_i^{abc}(x,y,z)=-\left(\epsilon_{ij}\partial_j
\left[\frac{1}{D\partial}\right]^{ad}\right)(x,u)
f^{dfe}\epsilon_{kl}
r_k^{fb}(u,y)r_l^{ec}(u,z)\, .
\end{equation}
Here again summation over the repeated color indices and integration
over the transverse coordinate $u$ on the right hand side is
understood.

Equations~(\ref{r}) and~(\ref{p}) give the complete solution for $r$
and $p$ in terms of the field $\alpha$.  The set of equations
Eq.~(\ref{correlf}) together with Eq.~(\ref{indfield}) and
Eqs.~(\ref{r}),~(\ref{p}) directly govern the evolution of the
correlators of chromoelectric field.

\begin{figure}[hbt]
\epsfysize=4cm
\begin{center}
\epsfysize=2.7cm
\begin{minipage}{4cm}
\epsfbox{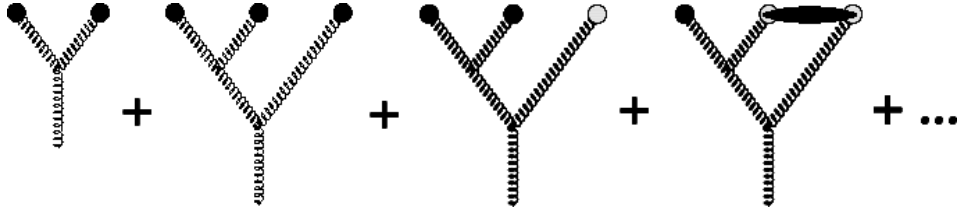}
\end{minipage}
\end{center}
\caption{\label{fig:one} \small \itshape 
  The diagrams contributing to the chromoelectric field at order
  $\alpha_s$. The full circles denote the background charge density
  $\rho$. The empty circles denote the average of the charge density
  induced by the fluctuations $<\delta\rho>=\sigma$. The black bars
  denote contractions corresponding to the mean fluctuation of the
  induced charge density $<\delta\rho\delta\rho>=\chi$.}
\end{figure}

\begin{figure}[hbt]
\epsfysize=4cm
\begin{center}
\epsfysize=2.7cm
\begin{minipage}{4cm}
\epsfbox{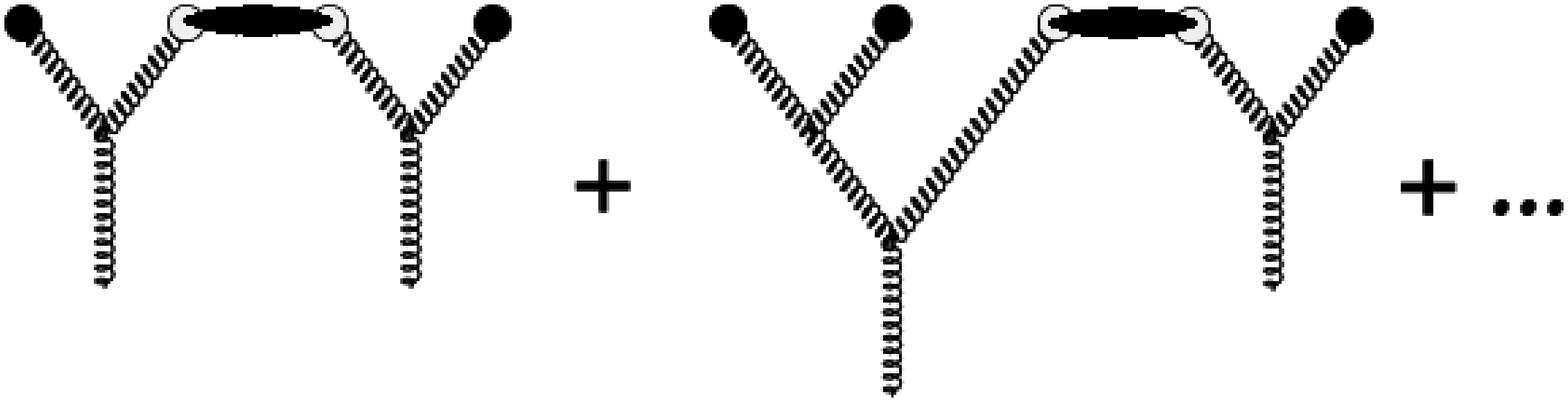}
\end{minipage}
\end{center}
\caption{\label{fig:two} \small \itshape 
 The average fluctuation of the induced chromoelectric field
$\chi^{ab}_{ij}$.
The notations are the same as in Fig.1.}
\end{figure}

We close this section by noting that the quantities $\chi^{ab}_{ij}$
and $\sigma^a_i$ have very simple physical meaning. As mentioned
above, the high momentum modes of the vector field which have been
integrated out in order to arrive at the evolution equation induce
extra color charge density $\delta\rho$.  The average value of this
induced density and its mean fluctuation appear in the evolution
equations for the correlators of charge density, as $\sigma^a$ and
$\chi^{ab}$.  Clearly the appearance of the induced color charge
density leads to the change in the value of the chromoelectric field
through the solution of Eq.~(\ref{sol}) with $\rho+\delta\rho$ on the
right hand side.  Diagrammatically the new field is represented by the
sum of the tree level diagrams with $\rho+\delta\rho$ as the source.
Those are depicted in Fig.~\ref{fig:one}.
It is clear that $\sigma^a_i$ is nothing but the average value of the
induced chromoelectric field.  Additionally, since the induced charge
density $\delta\rho$ contains components with frequencies $p^-$ higher
than those of the input charge density $\rho$, upon time averaging the
induced chromoelectric field is characterized also by mean fluctuation
$<\delta\alpha^a_i\delta\alpha^b_j>=\chi^{ab}_{ij}$. This again has
the simple diagrammatic representation shown in
Fig~\ref{fig:two}.\footnote{Given that Eq.~(\ref{final}) contains the
  average value and the mean fluctuation turn as the only relevant
  characteristics of the induced density $\delta\rho$ this naturally
  carries over to the the induced field $\delta\alpha$ to leading
  order in the coupling constant $\alpha_s$.}
The explicit expressions Eqs.~(\ref{indfield}),~(\ref{r} and~(\ref{p})
are equivalent to resumming all the diagrams of the type indicated in
Fig.~\ref{fig:one} and Fig.~\ref{fig:two}.  They clearly exhibit the
mechanism of the evolution of the field correlators in our approach.

\section{The Doubly Logarithmic Regime}

We would now like to consider the evolution in the doubly logarithmic
approximation.  By this we mean that in every step of the evolution
not only the longitudinal momenta are lowered but in addition, the
transverse momenta are required to grow. That is of course just the
QCD parton model picture that the "valence" partons have transverse
momenta on the typical hadronic scale while the high transverse
momentum partonic components of a hadron exist only as fluctuations at
short time scales. As discussed in \cite{moap} increasing the
frequency in the evolution is equivalent to transverse momentum
ordering in addition to the longitudinal momentum ordering.  This
scheme is therefore equivalent to the Born--Oppenheimer approximation
in a system with a continuum of time scales.

Technically, in our approach the doubly logarithmic regime is singled
out by restricting the transverse momenta in the fluctuation field
which is being integrated over to be smaller than the transverse
momentum of the observable that is being calculated, but larger than
the transverse momentum of the background field.  Practically this
just means that the background field throughout the calculation is
considered to be independent of $x_\perp$.  The results are then
easily obtained by taking the leading order of the transverse
derivative expansion of the general expressions for $\chi$ and
$\sigma$ given in \cite{moap}.  The result in this limit simplifies
very much for two main reasons.  First, since there is ordering in all
momentum components in the evolution it is clear that virtual diagrams
can not give a non-vanishing contribution.  This is simply the result
of momentum conservation in the $x_\perp$ independent background.
Therefore we have immediately
\begin{equation}
\sigma^a=\sigma^a_i=0 \, .
\end{equation}

The calculation of the real part also simplifies.  This is because the
two components of the chromoelectric field commute in this limit
\begin{equation}
[\alpha_i,\alpha_j]=0
\label{commut}
\end{equation}
This is the consequence of the first equation in Eq.~(\ref{sol})
for constant $\alpha_i$. This means that $\alpha_i$ can be treated as
constant numbers.  Using the explicit expressions from \cite{moap} for
$\chi$ we obtain in the double logarithmic limit
\begin{equation}
\chi^{ab}(x,y)=-4\left[\alpha^2
  <x|\frac{(\partial D)^2}{D^2(\partial^2+2\alpha^2)}|y>
  \right]^{ab}
\end{equation}
and
\begin{equation}
\label{dlchi}
\chi^{ab}_{ij}(x,y)=4\left[\alpha^2
<x|\frac{D_iD_j}{D^2(\partial^2+2\alpha^2)}|y>\right]^{ab}\, .
\end{equation}

Let us first check that our expressions in the limit of weak field do
indeed reproduce the DGLAP evolution in the double logarithmic regime,
as claimed.  To this end let us consider the evolution of the two
point function\footnote{In this equation as well as in the rest of the
  paper we use the bra and ket notation $<|>$ to denote both the
  averaging over the hadron state (averaging over $\rho$ in
  Eq.~(\ref{action})) of the function of $\alpha_i$'s and the
  configuration space matrix element of the powers of derivatives
  $\partial_i$.} .
\begin{equation}
\frac{d}{d \zeta}<\alpha_i^a(x)\alpha_j^b(y)>=4\alpha_s\left[<x|\alpha^2 
\frac{D_iD_j}{D^2(\partial^2+2\alpha^2)}|y>\right]^{ab}
\label{twop}
\end{equation}

At weak fields the $2\alpha^2$ term in the denominator of
Eq.~(\ref{twop}) should be dropped and the covariant derivatives
become simple derivatives.  Taking trace of the evolution equation
over both, color and transverse indices we obtain
\begin{equation}
\frac{d}{d \zeta}<\alpha_i^a(x)\alpha_i^a(y)>=-4\alpha_sN_c<\alpha_i^a
\alpha_i^a>
<x|\frac{1}{\partial^2}|y>
\label{dglap?}
\end{equation}

The gluon distribution is related to the field correlator in the
following way \cite{moap}
\begin{eqnarray}
xG(Q)&=&(2\pi)^2\int d^2b<F^{+i}(b,xp^+,x^+)F^{+i}(b, xp^+,x^+)>_Q
\nonumber\\
&=&
\frac{1} {4\pi \alpha_s}\int d^2b\int_{0}^{Q^2}d^2k
\int d^2x e^{ikx}<\alpha_i^a(b+x/2)\alpha_i^a(b-x/2)>
\label{distr}
\end{eqnarray}
Here $F^{+i}(b,k^+)=\int dx^-e^{ik^+x^-}F^{+i}(b,x^-)$ and the
subscript $Q$ means that the transverse coordinates in the
chromoelectric fields operators are equal with accuracy $1/Q$.  This
also gives
\begin{equation}
\frac{\partial}{\partial \ln Q^2}[xG(Q)]=\frac{1}{4\alpha_s} Q^2\int d^2b
\int d^2x e^{iQx}<\alpha_i^a(b+x/2)\alpha_i^a(b-x/2)>
\end{equation}
The factor $\frac{1}{4\pi\alpha_s}$ appears in these expressions since
the vector potential $\alpha_i$ differs in normalization from the
chromoelectric field by a factor $1/g$, see Eq.~(\ref{chrom}).  Taking
the Fourier transform of Eq.~(\ref{dglap?}) and integrating over the
impact parameter $b_i$ we obtain
\begin{equation}
\frac{\partial^2}{\partial \zeta\partial \ln
Q^2}[xG(Q)]=\frac{\alpha_sN_c}
{\pi} xG(Q)
\label{dglap}
\end{equation}
which is precisely the double logarithmic approximation to the DGLAP
equation.

The following comment is due here.  The right hand side of
Eq.~(\ref{dglap?}) contains the expression $<\alpha_i^a\alpha_i^a>$.
In the leading order of the derivative expansion $\alpha_i$ does not
depend on the transverse coordinate, and this expression of course is
also constant.  If one allows slow variation of the field (slow in the
sense that the scale of the spatial variation should be much larger than
$1/Q$) this obviously should be understood as
$<\alpha_i^a(b)\alpha_i^a(b)>$ where the impact parameter $b=(x+y)/2$
is the center of mass coordinate. Of course since the spatial
resolution is $1/Q$, the coordinate of the field is only defined to
this accuracy. The more exact specification can be only given beyond
the leading order of the derivative expansion and is therefore beyond
the approximation we are discussing here. Another way of saying this
is that the fields that contribute to the right hand side of the
evolution equation themselves contain only transverse momentum
components smaller than $Q$.  We therefore have to understand
$<\alpha_i^a\alpha_i^a>$ on the right hand side of Eq.~(\ref{dglap?})
as
\begin{equation}
<\alpha_i^a(b)\alpha_i^a(b)>=\int_{0}^{Q^2} \frac{d^2k}{4\pi^2}
\int d^2xe^{ikx}<\alpha^a_i(b-x/2) \alpha^a_i(b+x/2)>
\end{equation}
This is precisely how $G(Q)$ arises in the right hand side of
Eq.~(\ref{dglap}).  In this approximation therefore the dependence on
$Q$ and $b$ in Eq.~(\ref{dglap?})  factorizes and the products of the
fields are defined with the transverse cutoff $1/Q$.  This is a
general feature of the leading order of derivative expansion and is
not specific in any sense to the weak field limit.

We note that expanding the right hand side of Eq.~(\ref{twop}) to
second order in the field intensity and taking trace over color and
Lorentz indices we obtain the first nonlinear correction to the DGLAP
equation
\begin{equation}
\frac{\partial^2}{\partial \zeta \partial \ln
Q^2}xG(Q)=\frac{\alpha_sN_c}
{\pi} xG(Q)-{2\over Q^2}\int d^2b{\rm tr}<(\alpha(b)^2)^2>
\label{mq}
\end{equation}
This is precisely the contribution of the twist four operator to the
evolution of gluon density as calculated by Mueller and Qiu \cite{mq}.
It is probably worthwhile noting that Eq.~(\ref{mq}) should be
compared not with the nonlinear GLR equation, which is expressed in
terms of the distribution function, but rather with Eq.~(25) of
\cite{mq} which contains the contribution of twist four operators to
the evolution of the gluon distribution. It is straightforward to
check that taking into account the normalization of the operator
$G^{(2)}$ in \cite{mq} our Eq.~(\ref{mq}) is indeed equivalent to
Eq.~(25) of \cite{mq}. The nonlinear GLR equation arises as an
approximation to this equation if one assumes factorization of the
four gluon operator into the square of the gluon distribution in a
very particular manner.  Assuming this factorization Eq.~(\ref{mq})
would indeed lead to the GLR equation with the same coefficient as
given in \cite{mq}.

Equations for the higher correlators can be analyzed in a similar
fashion in the weak field limit. They all reduce to homogeneous
evolution equations and yield simple expressions for the anomalous
dimensions of the higher twist operators.  Our main interest here,
however, is to explore the opposite situation, namely when the fields
and partonic densities become large and this is what we will do in the
rest of this paper.

The most prominent feature of equation~(\ref{twop}) is the appearance
of the factor $Q^2-2\alpha^2$ in the denominator. The field strength
$-2\alpha^2$ therefore plays the role of a dynamically generated mass
in the evolution at high density\footnote{Note that the matrix
  $\alpha$ as we have defined it is anti hermitian. The eigenvalues of
  $\alpha^2$ are therefore negative.}.  This is not entirely
unexpected, but it is gratifying to see the emergence of this
dynamical mass in such a straightforward fashion in our approach.
Clearly the effect of this mass is to slow down the evolution.  In
fact it is easy to see that when the density becomes large this ``mass
term'' dominates the evolution and leads to unitarization.

Let us consider the evolution of the distribution function in the high
density -strong field limit $-\alpha^2>Q^2$. Neglecting $Q^2$ relative
to $\alpha^2$ in the denominator in Eq.~(\ref{twop})\footnote{This
  must be done with care, since the matrix $\alpha^2$ has zero
  eigenvalues. See comment following Eq.~(\ref{unit}).}  we obtain
\begin{equation}
\frac{\partial^2}{\partial \zeta\partial \ln Q^2}xG(Q)=
\frac{N_c(N_c-1)}{2} Q^2S
\label{unit}
\end{equation}
Here $S$ is the area of the hadron within which the density is high.
It appears due to the integration over the impact parameter. The
origin of the color factor $N_c(N_c-1)$ is the following.  If the
matrix $\alpha^2$ had no zero eigenvalues, this factor would be just
$N_c^2-1$ - the trace of the unit matrix in the adjoint
representation.  However the matrices $\alpha_i$ as defined in
Eq.~(\ref{def}) necessarily have zero eigenvalues. The number of these
zero eigenvalues is generically $N_c-1$. Since
$[\alpha_i,\alpha_j]=0$, the same applies to the matrix $\alpha^2$.
The contribution to the trace therefore comes only from the subspace
spanned by the eigenvectors which correspond to nonzero eigenvalues.
The dimensionality of this subspace is $N_c^2-1-(N_c-1)=N_c(N_c-1)$.

Integrating Eq.~(\ref{unit}) we obtain
\begin{equation}
xG(Q^2)=\frac{N_c(N_c-1)}{2} Q^2S\ln(1/x)+G_0
\label{unit1}
\end{equation}

This result is very interesting.  First it shows a characteristic
feature expected of unitarization~-~the growth with energy is only
logarithmic rather than power like as in the case of a pomeron, or
$\exp\sqrt{\log s}$ as in the case of the doubly logarithmic DGLAP
evolution.  On the other hand the gluon density does not completely
saturate but keeps growing logarithmically even at high densities.
Similar ``partial saturation'' was found in the analysis of
\cite{muller2}.  and also in the solution of a nonlinear evolution
equation in \cite{levin4}.

Identifying naively the gluon distribution with the total DIS cross
section for a probe that couples directly to gluons\footnote{This
  naive identification has to be taken with a grain of salt.  For
  large values of the gluon density Eq.~(\ref{unit1}) the leading
  twist relation between the cross section and the number of gluons is
  not valid. One then has to consider the contribution of higher twist
  operators directly in the cross section as well as in the evolution
  of the gluon density. The following argument should therefore be
  understood only in illustrative sense.}  one would get a logarithmic
growth of the cross section on top of the factor of the geometric area
of the hadron. This is perfectly compatible with the Froissart bound
and in fact grows slower than the allowed square of the logarithm if
the effective area $S$ in Eq.~(\ref{unit1}) grows slower than
logarithmically.  On the other hand the prefactor depends strongly on
the number of colors and is large at large $N_c$. This is not what one
expects generically from the unitarized cross section.  Once the
geometric size has been factored out one does not expect any other
numerical factors. The factor $N_c^2$ is of a purely perturbative
origin.  It counts the number of perturbative degrees of freedom -
gluons that take part in the scattering. This is in accord with the
fact that our approach is indeed perturbative and we therefore can
only detect the perturbative mechanism for unitarization which should
be operative in the region where the scattering is mainly
perturbative. Apart from the perturbative component which is described
by our present approach there is also a nonperturbative, low
transverse momentum one of the soft pomeron type (see e.g. the recent
analysis in \cite{dl}) which is not governed by the evolution we
consider here. At very low values of $x$ it must become leading and
eventually the perturbative behavior of
Eqs.~(\ref{unit}),~(\ref{unit1}) will cross over into the real
asymptotics which may behave as a square of the logarithm.

\section{The Gaussian model.}

Eq.~(\ref{unit1}) represents the asymptotics of the distribution at
large values of the density. At small $x$ one expects the density to
be large close to the center of a hadron, at small values of the
impact parameter, but not near the boundary.  Eq.~(\ref{unit})
therefore can not be quite correct even at small $x$.  To get a more
realistic picture one has to allow for a possibility of a more rapid
evolution in the peripheral regions.

Generically the system of equations Eq.~(\ref{correlf}) is a coupled
system of equations for infinite number of the multi gluon correlation
functions and as such is nontrivial even in the doubly logarithmic
limit. We would like to simplify this system so that it becomes more
tractable.  One straightforward possibility is to consider a Gaussian
truncation of the original system.  What we mean by this is to assume
standard Gaussian factorization of multi particle correlators of the type
\begin{equation}
<\alpha^{2n}>\sim(2n-1)!!<\alpha^2>^n 
\end{equation}
This approximation is in the spirit of the GLR model 
\cite{glr}.

Technically this means that we take the statistical weight used for
calculation of the averages on the right hand side of the evolution
equation as a Gaussian function of the fields.  One should of course
take care to ensure that this Gaussian weight is consistent with the
symmetries of the system and also with the fact that the two spatial
components of the chromoelectric field are not completely independent
but rather commute with each other, see Eq.~(\ref{commut}).  To ensure
this, let us explicitly solve the constraint commutativity imposes on
the matrices $\alpha_i$. The quantities to be averaged all being
invariant under global color rotations implies they can only depend on
the eigenvalues of the $\alpha_i$. Since the $\alpha_i$ commute, they
can be diagonalized simultaneously. We can therefore take the
following explicit representation\footnote{The most general matrix
  $\alpha_i$ has the form $iU\sum_A t_i^AT^AU^\dagger$, with $U$ an
  adjoint representation $SU(N)$ matrix. However, as noted before any
  $SU(N)$ invariant quantity does not depend on $U$ and we therefore
  omit it in all the following formulae.}
\begin{equation}
\alpha_i=i\sum_A t_i^AT^A
\end{equation}
Here $T^A$ are the diagonal Cartan subalgebra generators of the SU(N) group
in the adjoint representation, $t_i^A$ are real coefficients
and the index $A$ takes values from 1
to $N_c-1$. 
The coefficients $t_i^A$ are completely independent and the Gaussian
weight for the averaging can be taken as
\begin{equation}
\label{gaussian}
<...>=\left(\frac{(N_c-1)}{\alpha_s xg}\right)^{N-1}
\int \Pi_{i,A}dt_i^A
\ ...\ \exp\left\{-\frac{\pi(N_c-1)}{\alpha_sxg}t_i^At_i^A\right\}
\end{equation}
The meaning of the parameter $g$ is made clear by calculating the
glue--glue correlator
\begin{equation}
<\alpha_i^a\alpha_i^a>=\frac{\alpha_s}{\pi}xg \, .
\end{equation}
The quantity $g$ is therefore the local partonic density. Just as the
field itself it is of course a slowly varying function of the impact
parameter, and is related to the gluon distribution by
\begin{equation}
G(x,Q^2)=\int d^2b_\perp g(x, b_\perp, Q^2)
\end{equation}

Fourier transforming Eq.~(\ref{twop}) and taking trace we get for the
right hand side
\begin{equation}
4\alpha_s\sum_{n=1}^{N_c^2-1}
  <\frac{t_i^At_i^Br_n^Ar_n^B}{Q^2+2t_j^Ct_j^D
    r_n^Cr_n^D}>
\label{rhs}
\end{equation}
where $r_n^A$ is the $n$-th eigenvalue of the $A$-th Cartan subalgebra
generator in the adjoint representation. The $N_c-1$ dimensional
vectors ${\bf r}_n$ are the root vectors of $SU(N)$ and have the
following important properties
\begin{eqnarray}
{\bf r}_n^2&=&0, \ \ \ \ n=1,...,N_c-1 \\
{\bf r}_n^2&=&1, \ \ \ \ n=N_c,...,N_c^2-1\nonumber
\end{eqnarray}
Therefore only the terms $n=N_c,..., N_c^2-1$ contribute to the sum.
In each one of these non vanishing terms one can perform an orthogonal
rotation of the coefficients $t_i^A\rightarrow s_i^A$
such that
\begin{equation}
s_i^1=t_i^Ar_n^A
\end{equation}
The possibility of such an orthogonal rotation is assured by the fact
that the root vectors are properly normalized.  The expression in
Eq.~(\ref{rhs}) then simplifies to
\begin{eqnarray}
\lefteqn{\frac{4N_c(N_c-1)^2}{xg}
\int ds_i\exp\left\{-\frac{\pi(N_c-1)}{\alpha_sxg}
  \ s_i^2\right\}
\frac{s_i^2}{Q^2+2s_i^2}
} 
\\ &= &
\frac{\pi N_c(N_c-1)^2Q^2}{xg}\int_0^\infty d s
\left[\frac{s}{1+s}\right]\exp\left\{
  -\frac{\pi(N_c-1)Q^2}{2\alpha_sxg}\ s\right\}
\nonumber
\end{eqnarray}
The evolution equation then becomes
\begin{eqnarray}
\label{model}
\lefteqn{
\frac{\partial^2}{\partial \zeta \partial \ln Q^2}xg(x,Q,b_\perp)
}
\\ & = &
\frac{N_c(N_c-1)}{2}\ Q^2\bigg[1+
\frac{\pi (N_c-1)Q^2}{2\alpha_sxg}
\exp\left\{\frac{\pi(N_c-1)Q^2}{2\alpha_sxg}\right\}
{\rm Ei}\left(-\frac{\pi(N_c-1)Q^2}{2\alpha_sxg}\right)
\bigg]
\nonumber
\end{eqnarray}
where ${\rm Ei}(x)$ is the integral exponential function defined as
\begin{equation}
{\rm Ei}(x)=-\int_{-x}^\infty\! dt\ \frac{e^{-t}}{t},\ \ \ \ x<0
\end{equation}
 
We stress that the factorization assumed in Eq.~(\ref{gaussian}) is
motivated only to the extent that it leads to a closed evolution
equation for the gluon density.  One can think of it as a mean field
approximation for the true weight function for the averaging over
fields $\alpha_i$.  One certainly expects this approximation to be
valid at small $xg$ where it is equivalent to the steepest descent
integration over $\alpha_i$'s. Additionally at large $xg$ the
asymptotics of Eq.~(\ref{model}) reduces to the exact asymptotics of
the original equation~(\ref{twop}) and is independent of the form of
the weight function. It is therefore reasonable to expect that this
simple model provides a sensible interpolation of the evolution between
the weak and strong density regimes.

Note that Eq.~(\ref{model}) governs the evolution of the gluon density
$xg(b)$ and is local in the impact parameter space.  It therefore
contains more information on the structure of the hadron state than
just the gluon distribution $xG(x,Q^2)$.  To determine the evolution
of the gluon distribution $xG(x,Q^2)$ one has to integrate the
solution of Eq.~(\ref{model}) over the impact parameter $b$.  To write
down a closed equation directly for $xG$ one would have once again to
resort to modeling. This time one needs a model of the gluon density
as a function of the impact parameter.  Models of this kind have
indeed been used in the literature. One usually assumes a simple
factorizeable structure \cite{levin4} $g(x,Q^2,b)=G(x,Q^2)S^{-1}(b)$.
The resulting equation for $G$ then depends on the form of $S(b)$ one
takes in this ansatz. For example for a simple homogeneous disk model
$S^{-1}(b)=(\pi R^2)^{-1}\theta(R^2-b^2)$, Eq.~(\ref{model}) with the
substitution $Q^2\rightarrow Q^2\pi R^2$ governs the evolution of
$xG$.  More commonly one uses the Gaussian type distribution
$S^{-1}(b)=(\pi R^2)^{-1}e^{-{b^2\over 2R^2}}$.  Any specific ansatz
of this type introduces a new phenomenological parameter $R$ into the
equation which has the meaning of an effective radius of the gluon
distribution in the hadron. The same kind of parameter appears in the
GLR equation \cite{glr,mq}.

It would be interesting to study equation~(\ref{model}) numerically to
find out how the effective area in which the partonic density is high
($S$ in Eqs.~(\ref{unit}),~(\ref{unit1})) depends on the rapidity.
This we intend to do in a future publication.

\section{Discussion}

We have considered the doubly logarithmic limit of the low $x$
nonlinear evolution equations derived in \cite{moap}.  The most
salient feature in this regime is the appearance of a dynamically
induced mass which leads to the slowdown of the evolution as the
partonic density becomes larger. In particular we have shown that at
small impact parameters the gluon density grows as a logarithm of
energy rather than following the much faster growth pattern predicted
by DGLAP evolution.  To study the crossover between these two regimes
in detail one has to analyze the set of equations~(\ref{correlf}) for
the evolution of multi gluon correlators. This is still a formidable
problem, although in the doubly logarithmic limit the equations are
much simpler than the general evolution equations of \cite{moap}.

We have discussed an approximation in which a closed equation
Eq.~(\ref{model}) for the gluon density arises. This approximation
imposes Gaussian factorization of multi particle operators in terms of
the gluon density and is similar in spirit to the approximation used
to write down the GLR equation in \cite{glr} and \cite{mq}.

We note that the local equation for the gluon density
Eq.~(\ref{model}) is qualitatively similar to the equation suggested
in \cite{levin4}.  The authors of \cite{levin4} suggested the
following equation on the basis of reinterpretation of the Glauber
formula for multiple scattering of \cite{muller2}
\begin{equation}
\label{agl}
{\partial^2\over \partial \zeta\partial \ln Q^2}xg=
{2 Q^2\over \pi^3}
\left\{1-e^{-{N_c\alpha_s\pi^2\over 2Q^2}xg}\right\}
\end{equation}

Although the functional form of this equation looks somewhat different
from our Eq.~(\ref{model}) the main features are the same\footnote{The
  normalization of the gluon density in \cite{levin4} is slightly
  different than the one we use throughout this paper. This is the
  origin of the seemingly different factors of $\pi$ and $N$ in
  Eq.~(\ref{agl}) and Eq.~(\ref{model}). In fact rescaling $g$ in
  Eq.~(\ref{agl}) by ${4\over\pi^3(N-1)}$ brings the two equations
  into correspondence in the sense that both, the high density limit
  and the expansion parameter at low density become the same in the
  two expressions.}.  Both reduce to the DGLAP equation in the limit
of small density. Asymptotically at low $x$, where the density is high
the solution to both equations behaves as $Q^2{\rm ln}1/x$.

There are however also important differences between the two
equations.  At large densities the correction to the asymptotic
solution of Eq.~(\ref{agl}) is exponentially small. On the other hand
expanding Eq.~(\ref{model}) to order $1/xg$ we find
\begin{equation}
\frac{\partial^2}{\partial \zeta \partial \ln Q^2}xg(x,Q,b)=
\frac{N_c(N_c-1)}{2}\ Q^2\left[1-\frac{\pi (N_c-1)Q^2}{2\alpha_sxg}
\ln {2\alpha_sxg\over \pi (N_c-1)Q^2}\right]
\end{equation}
So the correction here is a power enhanced by a logarithmic factor.
The approach to the asymptotics is therefore much slower than in
Eq.~(\ref{agl}). Since the correction term is negative it is clear
that the solution to Eq.~(\ref{model}) has a slower growth at low $x$.
That is to say, at low densities solutions of both Eq.~(\ref{agl}) and
Eq.~(\ref{model}) start growing according to the double logarithmic
DGLAP equation $xg\propto e^{\sqrt{\bar\alpha_s\ln 1/x}}$ (neglecting
the running of $\alpha_s$). This growth however is tempered faster in
the solution of Eq.~(\ref{model}) than in Eq.~(\ref{agl}).  One could
wonder perhaps whether the existence of power corrections at large
density in Eq.~(\ref{model}) is an artifact of the Gaussian truncation
of the infinite set of equations Eq.~(\ref{correlf}). It is easy to
see that this is not the case. The reason is that the mechanism of
screening in our approach is the appearance of the effective mass
proportional to the intensity of the chromoelectric field in
Eq.~(\ref{dlchi}). This type of screening with necessity gives power
like preasymptotic corrections.  It would be very interesting to
understand better the relation between our present approach and the
approach of \cite{levin4}.

Finally we note that in a recent very interesting paper \cite{makhlin}
Makhlin and Surdutovich also considered the appearance of effective
mass in the context of nuclear collisions.  Their approach physically
is complementary to ours.  They consider the emission in the final
state in a nuclear collision. In this situation the large transverse
momentum modes are emitted first whereas the low transverse momenta
are then emitted and propagate not in the vacuum but in the background
of these hard modes. The propagation of soft modes in the preexisting
hard mode background is then characterized by an effective mass.  We
on the other hand consider fast, large transverse momentum
fluctuations in the hadron which develop on the background of a
slower, smaller transverse momentum fields. At large fields the
fluctuations are inhibited and this is the origin of the effective
mass in our approach.  It characterizes the emission of hard
fluctuations in the preexisting soft field background.

{\small \it \subsubsection*{ Acknowledgements}We thank G. Levin, A.
  Makhlin and G. Surdutovich for interesting discussions.J.J-M was
  supported by the Director, Office of Energy Research, Office of High
  Energy and Nuclear Physics Division of the Deparment of Energy,
  under contract No. DE-AC03-76SF00098 and DE-FG02-87ER40328.  A.K. is
  grateful to the Nuclear Theory group at LBL and the Theoretical
  Physics Institute, University of Minnesota for hospitality during
  April 1998 when part of this work was done.  A.K.  is supported by
  PPARC Advanced Fellowship.  A.L. is partially supported by Russian
  Fund for Basic Research under Grant 96-02-16210.  H.W. was supported
  by the EC TMR Program, contract ERB FMRX-CT96-0008. He is grateful
  to the Theoretical Physics Institute, University of Minnesota for
  their hospitality during March and April 1998 and wants to thank the
  Alexander von Humboldt Foundation for support during that stay.  }

\end{document}